\documentclass[twocolumn,amsmath,amssymb]{snp}
\usepackage{epsfig}
\usepackage{hyperref}
\usepackage{url}
\usepackage{graphicx}% Include figure files
\usepackage{dcolumn}% Align table columns on decimal point
\usepackage{bm}% bold math
\usepackage{graphicx,amssymb}
\usepackage{color}
\usepackage{float}
\newcommand{\pt}{$p_{ {\mathrm T} }$}
\newcommand{\y}{$y$}

\usepackage{multirow}
\usepackage{rotating}
\usepackage{lineno}
\usepackage{subfigure}
\begin{document}

%\title{\Large Suppression of J/$\mathbf{\psi}$ and $\mathbf{\psi}$(2S) production in \mbox{p-Pb} collisions at $\mathbf{\sqrt{{\textit s}_{\rm NN}}}$~=~5.02 TeV}% Force line breaks with \\
\title{\Large Suppression of inclusive J/$\mathbf{\psi}$ and $\mathbf{\psi}$(2S) production in \mbox{p-Pb} collisions with ALICE at the LHC}% Force line breaks with \\

\author{\large Biswarup Paul (for the ALICE Collaboration)}
\email{biswarup.paul@cern.ch}
%\email{biswarup.paul@saha.ac.in}
%\author{\large Sukalyan Chattopadhyay (For ALICE Collaboration)\\}
\affiliation{Saha Institute of Nuclear Physics, Kolkata - 700064, India}
\maketitle

%\linenumbers
The ALICE Collaboration has studied both J/$\psi$ and $\psi$(2S) production in p-Pb collisions at $\sqrt{s_{\rm NN}} = 5.02$ TeV~\cite{jhep1402,arxiv1405,arnaldi}, through their dimuon decay channel, in the Muon Spectrometer which covers the pseudorapidity range $-$4 $\leq \eta \leq$ $-$2.5. The ALICE detector is described in detail in~\cite{jinst}. Data have been collected under two different configurations, inverting the direction of the p and Pb beams. In this way both forward ($2.03 \leq y_{\rm cms} \leq 3.53$) and backward ($-4.46 \leq y_{\rm cms} \leq -2.96$) centre of mass rapidities could be accessed, with the positive \y\ defined in the direction of the proton beam. The difference in the covered \y\ ranges reflects the shift of the centre of mass of the nucleon-nucleon collisions ($\Delta{y}_{\rm NN}$ = 0.465) with respect to the laboratory frame, induced by the different energies per nucleon of the colliding beams. The J/$\psi$ and $\psi$(2S) yields are extracted by fitting the dimuon invariant mass distributions with a superposition of signals and background shapes. For the signal, pseudo-Gaussian or Crystal Ball functions with asymmetric tails on both sides of the resonance peak are used, while for the background a Gaussian with a mass-dependent width or polynomial $\times$ exponential functions are adopted.

\begin{figure}[h]
\includegraphics[scale=0.33]{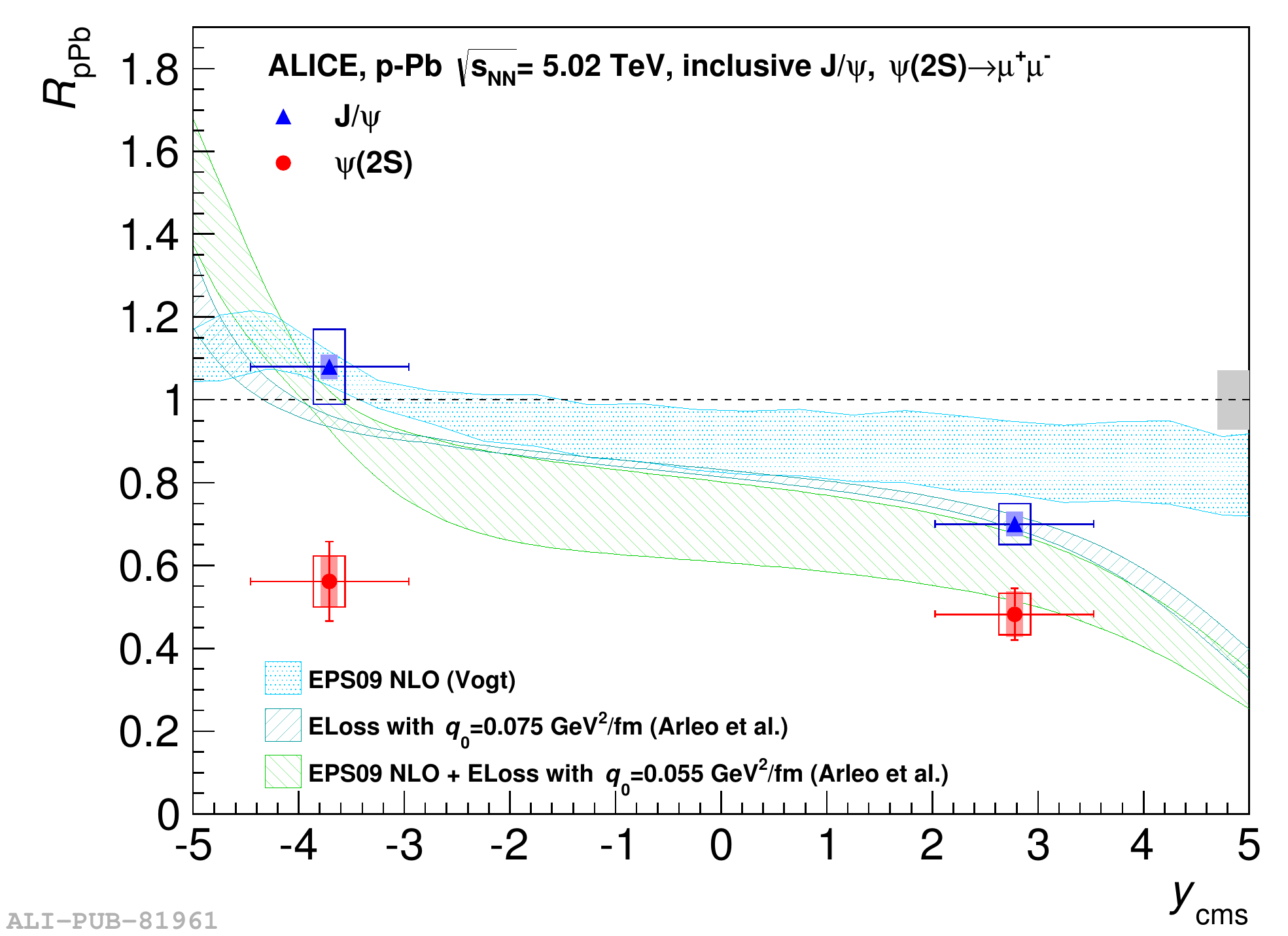}
\vspace{-3.00mm}
\caption{\label{fig1} J/$\psi$ and $\psi$(2S) $R_{\rm pPb}$ versus \y\ compared to theoretical models.}
\vspace{-3.00mm}
\end{figure}
The production cross section of $\psi$(2S) in p-Pb is compared to the J/$\psi$ one and to the corresponding quantities in pp collisions at $\sqrt{s}$ = 7 TeV (no LHC results are available at $\sqrt{s}$ = 5.02 TeV) using the double ratio [$\sigma_{\rm \psi(2S)}/\sigma_{\rm J/\psi}]_{\rm pPb}$/[$\sigma_{\rm \psi(2S)}/\sigma_{\rm J/\psi}]_{\rm pp}$~\cite{arxiv1405,arnaldi}. The nuclear modification factor of $\psi$(2S) is obtained by combining the J/$\psi$ $R_{\rm pPb}$~\cite{jhep1402} and the double ratio, as\\ $R_{\rm pPb}^{\rm \psi(2S)} =$$ R_{\rm pPb}^{\rm J/\psi}$$\times$$(\sigma_{\rm pPb}^{\rm \psi(2S)}/\sigma_{\rm pPb}^{\rm J/\psi})$$\times$$(\rm \sigma_{\rm pp}^{\rm J/\psi}/\sigma_{\rm pp}^{\rm \psi(2S)})$.\\ In Fig.~\ref{fig1}, $R_{\rm pPb}^{\psi(2S)}$ is compared with $R_{\rm pPb}^{J/\psi}$ and also with theoretical calculations based on nuclear shadowing~\cite{ijmp}, coherent energy loss or both~\cite{jhep1303}. The suppression of $\psi$(2S) production is much stronger than that of J/$\psi$ and reaches a factor of 2 with respect to pp. Since the kinematic distributions of gluons producing the J/$\psi$ or the $\psi$(2S) are rather similar and since the coherent energy loss does not depend on the final quantum numbers of the resonances, the same theoretical calculations hold for both J/$\psi$ and $\psi$(2S). Theoretical models predict \y\ dependence which are in reasonable agreement with the J/$\psi$ results but no model can describe the $\psi$(2S) data. These results show that other mechanisms must be invoked in order to describe the $\psi$(2S) suppression in p-Pb collisions.  

\begin{figure}
\includegraphics[scale=0.33]{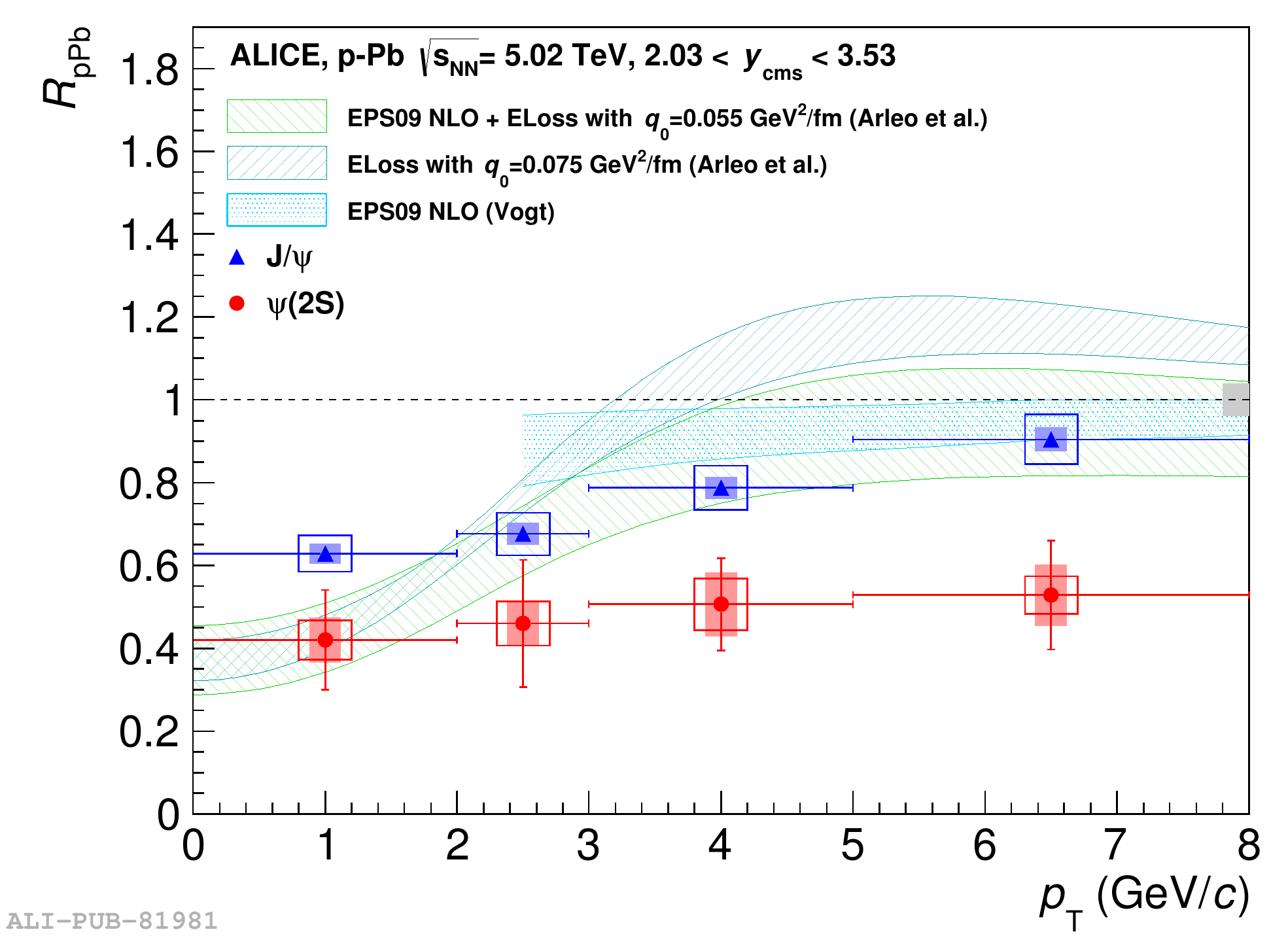}
\vspace{-3.00mm}
\caption{\label{fig2} \pt\ dependence of J/$\psi$ and $\psi$(2S) $R_{\rm pPb}$ compared to theoretical calculations in the forward \y\ region.}
\vspace{-3.00mm}
\end{figure}
\begin{figure}
\includegraphics[scale=0.33]{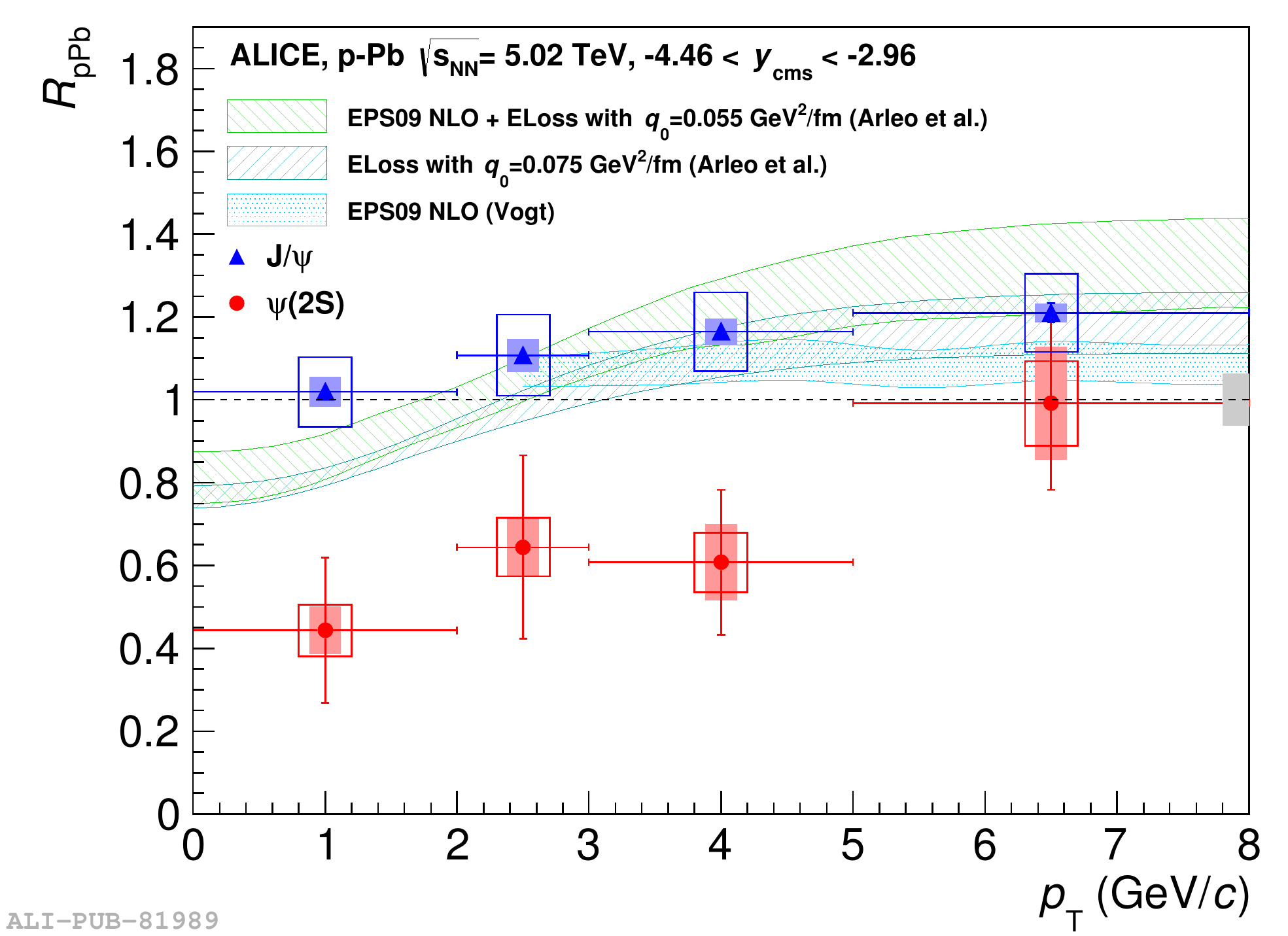}
\vspace{-3.00mm}
\caption{\label{fig3} Same as Fig~\ref{fig2} but in the backward \y\ region.}
\vspace{-3.00mm}
\end{figure}
\begin{figure}
\includegraphics[scale=0.33]{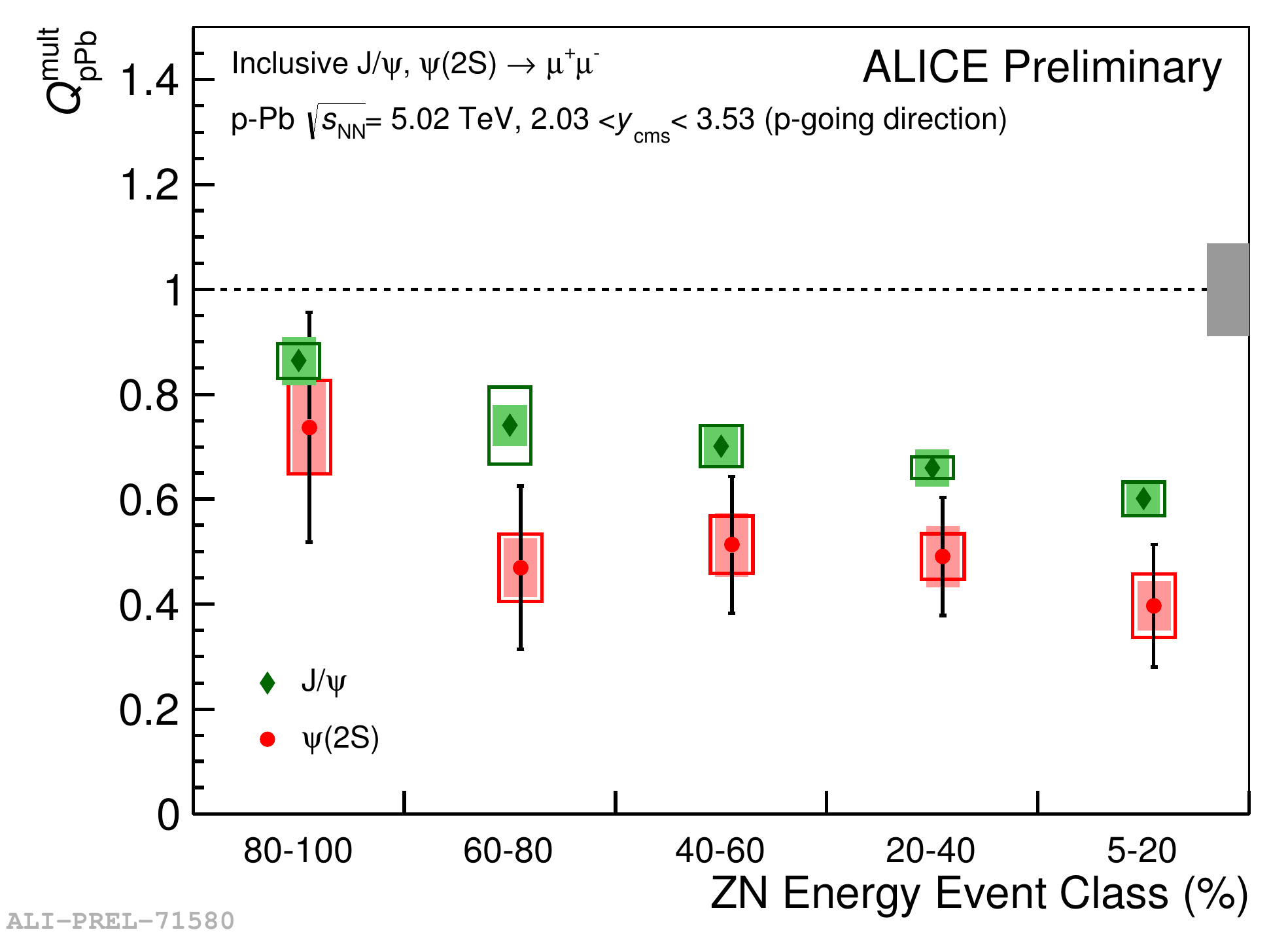}
\vspace{-3.00mm}
\caption{\label{fig4} J/$\psi$ and $\psi$(2S) $Q_{\rm pPb}$ versus event activity in the forward \y\ region.}
\vspace{-3.00mm}
\end{figure}
\begin{figure}
\includegraphics[scale=0.33]{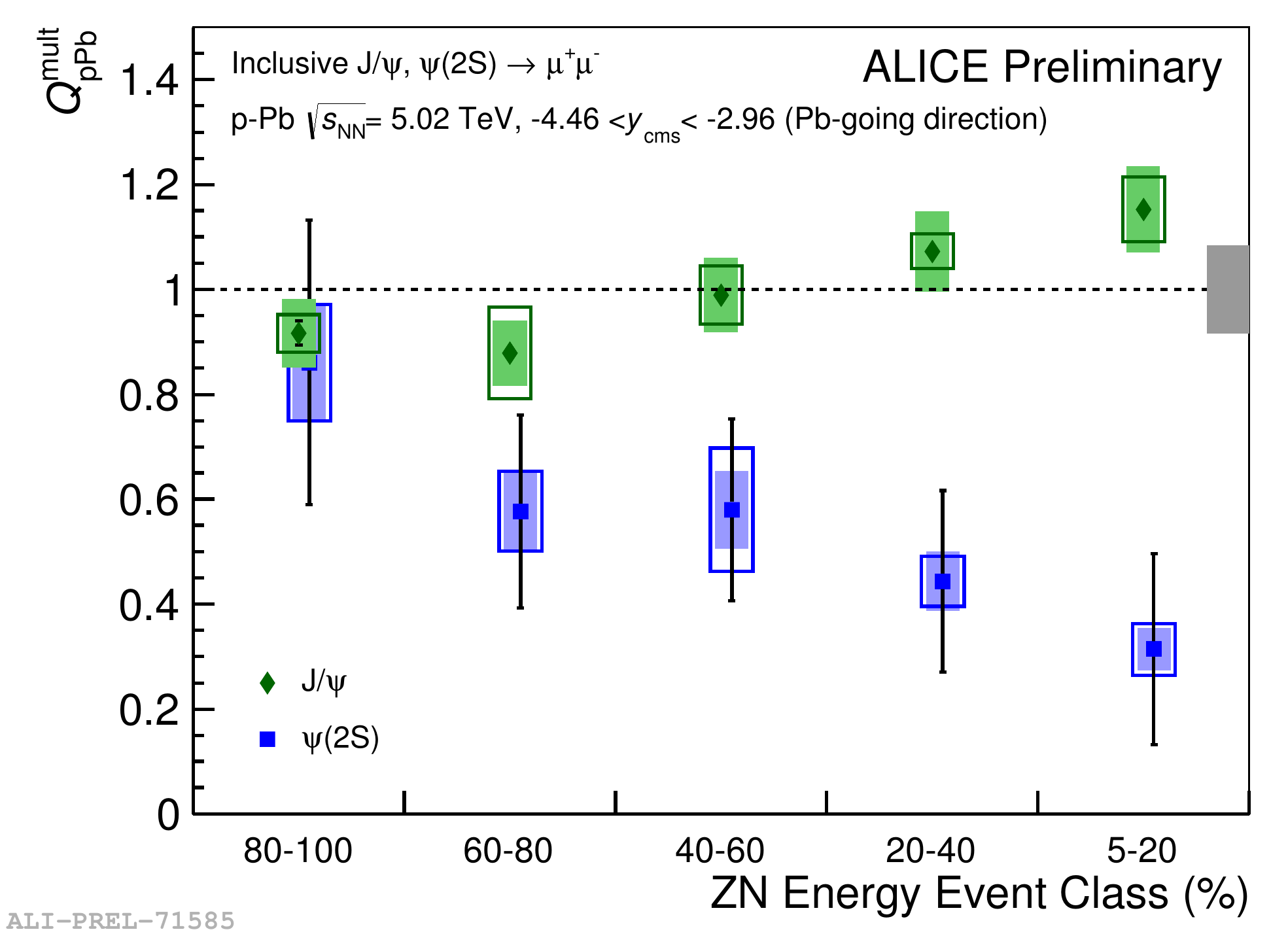}
\vspace{-3.00mm}
\caption{\label{fig5} Same as Fig~\ref{fig4} but in the backward \y\ region.}
\vspace{-3.00mm}
\end{figure}
The $R_{\rm pPb}$ is also computed as a function of $p_{\rm T}$ both at backward and forward \y\ and the results are shown in Fig.~\ref{fig2} and~\ref{fig3}, respectively. At both rapidities, the $R_{\rm pPb}^{\rm \psi(2S)}$ shows a strong suppression with a slightly more evident $p_{\rm T}$ dependence at backward-\y. The $\psi$(2S) is more suppressed than the J/$\psi$, as already observed for the $p_{\rm T}$-integrated result. Theoretical calculations are in fair agreement with the $R_{\rm pPb}^{\rm J/\psi}$ but clearly overestimate the $R_{\rm pPb}^{\rm \psi(2S)}$ behaviour.

Finally, the $\psi$(2S) production is studied as a function of the event activity both at backward and forward \y~\cite{arnaldi}, as shown in Fig.~\ref{fig4} and~\ref{fig5}, respectively. The event activity determination is described in details in Refs.~\cite{toia}. Since the centrality estimation in p-Pb collisions can be biased by the choice of the estimator, the nuclear modification factor is, in this case, named $Q_{\rm pPb}$~\cite{toia}. The $\psi$(2S) $Q_{\rm pPb}$ shows a strong suppression, which increases with increasing event activity, and is rather similar in both the forward and the backward \y\ regions. The J/$\psi$ $Q_{\rm pPb}$ shows a similar decreasing trend at forward-\y\ as a function of the event activity. On the contrary, the J/$\psi$ and $\psi$(2S) $Q_{\rm pPb}$ patterns observed at backward-\y\ are rather different, with the $\psi$(2S) significantly more suppressed for large event activity classes.

\end{document}